\input harvmac
\newcount\figno
\figno=0
\def\fig#1#2#3{
\par\begingroup\parindent=0pt\leftskip=1cm\rightskip=1cm\parindent=0pt
\baselineskip=11pt
\global\advance\figno by 1
\midinsert
\epsfxsize=#3
\centerline{\epsfbox{#2}}
\vskip 12pt
{\bf Fig. \the\figno:} #1\par
\endinsert\endgroup\par
}
\def\figlabel#1{\xdef#1{\the\figno}}
\def\encadremath#1{\vbox{\hrule\hbox{\vrule\kern8pt\vbox{\kern8pt
\hbox{$\displaystyle #1$}\kern8pt}
\kern8pt\vrule}\hrule}}

\overfullrule=0pt

%macros
%
\def\tilde{\widetilde}

\def\Z{{\bf Z}}
\def\C{{\bf C}}
\def\T{{\bf T}}
\def\S{{\bf S}}
\def\R{{\bf R}}

\font\zfont = cmss10 %scaled \magstep1

\def\bigone{\hbox{1\kern -.23em {\rm l}}}
\def\ZZ{\hbox{\zfont Z\kern-.4emZ}}

\Title{hep-th/9604030, IASSNS-HEP-96-29}
{\vbox{\centerline{Non-Perturbative Superpotentials}
\bigskip
\centerline{ In String Theory}}}
\smallskip
\centerline{Edward Witten\foot{Research supported in part
by NSF  Grant  PHY-9513835.}}
\smallskip
\centerline{\it School of Natural Sciences, Institute for Advanced Study}
\centerline{\it Olden Lane, Princeton, NJ 08540, USA}\bigskip

\medskip

\noindent
The non-perturbative superpotential can be effectively calculated
in $M$-theory compactification to three dimensions on a Calabi-Yau
four-fold $X$.  For certain $X$, the superpotential is identically
zero, while for other $X$, a non-perturbative superpotential is
generated.  Using $F$-theory, these results carry over to certain
Type IIB and heterotic string compactifications to four dimensions
with $N=1$ supersymmetry.    In the heterotic string case, the 
non-perturbative superpotential can be interpreted as coming
from space-time and world-sheet instantons; in many simple cases
contributions come only from finitely many values of the instanton
numbers.

\Date{April, 1996}
%text of paper

\newsec{Introduction}

Surprisingly much of the dynamics of supersymmetric field theories
and string theories has proved to be knowable,
leading one to wonder how much farther one can get 
using techniques that are more or less
already available.  In particular, one would very much like to obtain
non-perturbative information about the superpotential of $N=1$
compactifications to four dimensions.

In this paper, we take some steps in this direction.
We consider in section two the compactification of eleven-dimensional
$M$-theory to three dimensions on a manifold $X$ of $SU(4)$ holonomy.
This gives a model with $N=2$ supersymmetry in three dimensions,
which is roughly comparable to $N=1$ in four dimensions.
We argue that superpotentials in this model are generated {\it entirely}
by  instantons obtained by wrapping a five-brane over a complex
divisor $D$ in $X$.  Moreover, only very special $D$'s can contribute;
for a given $X$ one can effectively find all of the relevant $D$'s,
which in many simple examples are finite in number,
and obtain a fairly precise formula 
(which depends on one loop determinants that are hard to make
explicit) for the superpotential.  For many
$X$'s -- such as complete intersections in projective spaces -- there
are no $D$'s with the right properties, and the superpotential is
identically zero; supersymmetry is thus unbroken in these models.
Other simple examples generate non-trivial superpotentials.

Some of these models can be directly related to four-dimensional
models via $F$-theory \ref\vafa{C. Vafa, ``Evidence For
$F$-Theory,'' hepth/9602022.}.
If $X$ admits an elliptic fibration, that is if there is a holomorphic
map $X\to B$ with the generic fiber being an elliptic curve, then
$M$-theory on $X$ goes over, in a certain limit, to Type IIB superstring
theory on $B$, which is a {\it four-dimensional} theory with $N=1$
supersymmetry.  Thus, by taking suitable limits of the formulas of
section two, one gets, as we discuss in section three,
exact superpotentials for this class of four-dimensional
$N=1$ models.

\def\P{{\bf P}}
Moreover, if $B$ in turn is ``rationally ruled,'' that is if there is
a holomorphic 
fibration $B\to B'$ with the fibers being ${\bf P}^1$'s, then
Type IIB on $B$ is equivalent to the heterotic string on a Calabi-Yau
threefold $Z$ that is elliptically fibered over $B'$.  This conclusion
follows upon fiberwise application of the equivalence \vafa\
of the heterotic string on a two-torus with Type IIB on a two-sphere,
and (for $B$ of complex dimension two) has been
used to study the heterotic string on K3 \ref\morrisonvafa{D. Morrison
and C. Vafa, ``Compactifications Of $F$-Theory On Calabi-Yau
Threefolds, I, II,'' hepth/9602114,9603161.}.  
So, as we discuss in section four,
the results of section three imply exact,
non-perturbative formulas for the superpotentials in certain
Calabi-Yau compactifications of the heterotic string.
As might be expected, these superpotentials can be interpreted
as sums of contributions from world-sheet instantons and space-time
instantons.   Given the results mentioned above in connection with 
$M$-theory, it turns out that, in many simple cases, 
non-vanishing contributions to the world-sheet and space-time instanton 
numbers arise only for {\it finitely} many values of the instanton
numbers.

\nref\ewitten{E. Witten, ``Some Comments on String Dynamics,''
hepth/9507121.}
\nref\otherstrom{A. Strominger, ``Open $p$-Branes,'' hepth/9512059.}
\nref\hanany{O. Ganor and A. Hanany, ``Small Instantons And
Tensionless Strings,'' hepth/9602120.}
\nref\seibergwitten{N. Seiberg and E. Witten,
``Comments On String Dynamics In Six Dimensions,'' hepth/9603003.}
\nref\newduff{M. Duff, H. Lu, and C. N. Pope, ``Heterotic Phase
Transitions And Singularities Of The Gauge Dyonic String,'' 
hepth/9603037.}
\nref\fm{E. Witten, ``Phase Transitions In $M$-Theory And
$F$-Theory,'' hepth/9603150.}

Ironically, these non-perturbative results give new information
about the heterotic string  even at the {\it perturbative}
level.  It has long been known \ref\dsww{M. Dine, N.  Seiberg, X.-G. Wen,
and E. Witten, ``Non-Perturbative Effects On The String 
World-Sheet,'' Nucl. Phys. {\bf B278} (1986) 769, {\bf B289} (1987)
319.} that in principle
world-sheet instantons should generate a superpotential
that would spoil the conformal invariance of $(0,2)$ sigma models.  
Yet a concrete example in which this actually occurs has not been found.
Indeed, global holomorphy
can sometimes be used \ref\silver{E. Silverstein and E. Witten,
``Criteria For Conformal Invariance Of $(0,2)$ Models,''
Nucl. Phys. {\bf B444} (1995) 161.} 
to show
that cancellations must occur between different instantons of the
same instanton number, preventing the generation of a superpotential, 
and it has actually appeared difficult to see how such cancellations
could be avoided.  One mechanism for avoiding such cancellation
and actually generating a world-sheet instanton superpotential
will become clear in this paper.

The goal of understanding the superpotential is, of course,
to understand supersymmetry breaking and the vanishing of the cosmological
constant.  The main clue we get in that direction may be that
the special divisors from which the superpotential is generated
are often the ones whose collapse can sometimes
lead (at least in similar problems above four dimensions)
to novel infrared physics with,
roughly, tensionless strings \refs{\ewitten - \fm} and an abrupt
``end'' of the moduli space.  Since the vanishing of the cosmological
constant seems to defy understanding in terms of conventional
infrared physics, the fact that the mechanism (or at least the divisor)
that generates a superpotential may also generate novel infrared physics
may be encouraging.   

\newsec{$M$-Theory On A Calabi-Yau Four-Fold}

We consider compactification of $M$-theory from eleven to three
dimensions on a manifold $X$ of holonomy $SU(4)$. This gives a theory
with three-dimensional $N=2$ supersymmetry  (which
has a structure very similar to $N=1$ in four dimensions) in which there
are three kinds of supermultiplets
 that contain scalar fields:

(1) Fields derived from the complex structure of $X$ are the
scalar components of chiral multiplets.

(2) Fields obtained by integrating the three-form potential $C$
of eleven-dimensional supergravity over three-cycles in $X$ are
 likewise  scalar components of chiral multiplets.

\nref\siegel{W. Siegel, ``Gauge Spinor Superfield As A Scalar
Multiplet,'' Phys. Lett. {\bf 85B} (1979) 333.}
\nref\ogates{S. J. Gates, ``Super $p$-Form Gauge Superfields,''
Nucl. Phys. {\bf B184} (1981) 381. }
\nref\gates{S. J. Gates, M. Grisaru, M. Rocek, and W. Siegel,
{\it Superspace} (Benjamin-Cummings, 1983), pp. 181-197.}
\nref\othergates{H. Nishino and S. J. Gates, ``Chern-Simons
Theories With Supersymmetries In Three Dimensions,''
Int. J.  Mod. Phys. {\bf 8}
(1993) 3371.}
(3) There is also one slightly more exotic case analogous to
the ``linear multiplet'' in four dimensions \refs{\siegel - \gates}.
Let $r$ be the dimension of $H^2(X,\R)$ (which coincides
with the dimension of $H^{1,1}(X)$, since $H^{2,0}$ vanishes
for manifolds of holonomy $SU(N)$).  Then there are $r$
multiplets, known as three-dimensional linear multiplets, with the
following structure.  (Such multiplets have been discussed in
\othergates.)   The bosonic fields in such a multiplet
consist of one
real scalar that is obtained by integrating the Kahler form 
$\omega$ of $X$ over a two-cycle $E$,  along with a three-dimensional
vector field $A$ that is obtained by integrating the three-form
$C$ over $E$.  
Note that in three dimensions, a vector is dual to a scalar,
and  if one performs a duality transformation to convert $A$ into
a scalar $\phi$, one gets a conventional chiral supermultiplet with two
real scalar fields.  We will call the $\phi$'s obtained this
way ``dual scalars.''  

Now, there are no terms in  the superpotential that
are independent of the linear multiplets.  In fact,
such terms, being independent of the Kahler class of $X$, could
be computed by scaling up the metric of $X$; but as the metric
is scaled up, $M$-theory reduces to eleven-dimensional supergravity,
which has $\R^3\times X$ as an exact solution (since $X$ obeys the
Einstein equations), showing
that there is no superpotential in this limit.

At  first sight, it also seems impossible to have a superpotential
interaction that depends on the linear multiplets.  In fact,
it appears that the gauge field 
$A$ can only have derivative couplings, through
the gauge invariant field $F=dA$; in that case, the scalar $\phi$,
introduced by $d\phi=* F$, likewise only has derivative couplings,
so that a superpotential depending on $\phi$ is impossible.  

However, there is a fallacy in the last claim.  In
a situation  such as this, interactions
that are not invariant under $\phi\to \phi+{\rm constant}$ are indeed
absent perturbatively, but can
be generated \ref\polyakov{A. M. Polyakov, ``Quark Confinement And The
Topology Of Gauge Groups,'' Nucl. Phys. {\bf B120} (1977) 429.}
 by certain kinds of instanton, namely those that look like magnetic
monopoles for the $F$-field.  In other words, in the relevant instanton
field, $F$ has a non-zero integral over a large sphere at infinity
in $\R^3$, and therefore decays at infinity as $1/r^2$, with
$r$ the distance to the origin.  Dually, this means that
in the instanton field, $\phi$ falls off as $1/r$, corresponding
to the effects of a source at the origin that is linear in $\phi$
and not invariant under addition to $\phi$ of a constant.  The
interactions generated by such an instanton are proportional
to 
\eqn\jumbo{e^{-i\gamma \phi},} 
where the constant $\gamma$ is proportional
to the magnetic charge of the instanton.    

So far, we have not assumed supersymmetry.  If we do incorporate
three-dimensional $N=2$ supersymmetry, these interactions will 
be superpotential terms precisely
if  the instanton is  invariant under
two of the four supersymmetry charges  \ref\harvey{I. Affleck,
J. A. Harvey, and E. Witten, ``Instantons and (Super)Symmetry
Breaking In $2+1$ Dimensions,'' Nucl. Phys. {\bf B206} (1982) 413.}. 
This ensures that the instanton
generates a superpotential term $\int d^2\theta \,\,\dots$ and
not a generic coupling $\int d^4\theta \,\,\dots$.  

In the present context, $A$ originates as a mode of $C$, so the
requisite
instanton is a magnetic source for $C$.  In general, the magnetic source
(as reviewed in \ref\khuri{M. J. Duff and R. Khuri,
``String Solitons,'' Phys. Rept. {\bf 259} (1995) 213.}) 
is the eleven-dimensional
five-brane.  So roughly as in \ref\becker{K. Becker,
M. Becker, and A. Strominger, ``Five-Branes, Membranes, And
Non-Perturbative String Theory,'' Nucl .Phys. {\bf B456} (1995). } 
the relevant
instantons must be made by wrapping the six-dimensional world-volume
of the five-brane over a six-cycle $D$ in $X$, giving
what in three dimensions looks like an instanton.  This instanton
 is invariant under
some supersymmetry  precisely if  $D$ is a complex divisor, that is,
a complex submanifold of $X$.  We will actually see that $D$ must
obey an additional condition that eliminates most divisors.  

The amplitude of such an instanton is proportional to $e^{-V_D}$,
where $V_D$ is the volume of $D$, measured in units of the five-brane
tension.  There is an additional factor $e^{-i\phi_D}$, where as in
\jumbo, $\phi_D$ is a linear combination of the dual scalars.  
These factors thus combine
to a factor of $e^{-(V_D+i\phi_D)}$, strongly suggesting that
$V_D$ and $\phi_D$ are the real and imaginary parts of a chiral
supermultiplet.  
This can indeed be verified directly in eleven-dimensional
supergravity.  (Beyond the supergravity approximation, $\phi_D$ 
-- being dual to a vector, which is not subject to nonlinear
change of variable -- is naturally defined up to an additive constant
in the exact $M$-theory; that is not so for $V_D$, which in the
exact theory one can simply define to be the superpartner of $\phi_D$.)
The factor
$e^{-(V_D+i\phi_D)}$ thus has the holomorphy expected of a superpotential.

The $\phi_D$ dependence is here exactly fixed by the 
magnetic charge of the instanton, so the $V_D$ dependence is in turn
exactly fixed by holomorphy.  This then means that -- apart from
the known factor $e^{-(V_D+i\phi_D)}$ -- the superpotential generated
by the instanton is independent of the Kahler class of $X$, and so
can be computed by scaling up the metric of $X$.  The instanton
amplitude is computed by multiplying a classical factor
$e^{-(V_D+i\phi_D)}$ times a one-loop determinant of world-volume
fields -- which is invariant under scaling the metric of $X$.
Higher loop corrections to the world-volume computation would
be proportional to inverse powers of the Kahler class and so in
fact vanish by holomorphy.  The one-loop approximation to the
instanton amplitude is thus {\it exact},  for the purposes of
computing the superpotential.  This is a situation often
found, for somewhat analogous reasons, in computations of
superpotential generation by space-time \ref\ads{I. Affleck, M. Dine,
and N. Seiberg, ``Dynamical Supersymmetry Breaking In Four
Dimensions And Its Phenomenological Implications,''
Nucl. Phys. {\bf B256} (1985) 557.} or world-sheet \dsww\ instantons.  

The one-loop amplitude must also depend holomorphically on
the other variables -- and notably on the complex structure of
$X$, which enters via the complex structure of $D$.  It would
be of interest to analyze this dependence, which presumably
involves something similar to Ray-Singer analytic torsion
of complex manifolds.  What makes it difficult to immediately
identify the one-loop determinants in this problem
with anything known is that
the world-volume theory on $D$ is somewhat exotic, because of the
presence of a two-form with self-dual field strength.   

But one important property of the one-loop amplitude is easy
to extract: the contribution to the superpotential vanishes, because of an
anomaly, unless $D$ obeys a certain rather  strong condition.  This 
result will be obtained by matching instanton quantum numbers with
quantum numbers needed for a superpotential in a way roughly
familiar \ads\ from field theory analyses of instanton-generated
superpotentials, though there will be some unusual details in
the present case.  

In the analysis, I will assume that $D$ is smooth.
Since $D$ is of complex codimension one in $X$,
the normal bundle to $D$ in $X$  is a complex line bundle $N$.  The fact
that the canonical bundle of $X$ is trivial means that $N$ is isomorphic
to the canonical bundle $K_D$ of $D$.  Locally, near $D$, $X$ looks
like the total space of the normal bundle;   this approximation becomes
better as the metric is 
scaled up, and is exact for the world-volume theory
in the linearized approximation.  If $z$ is a local
coordinate in the normal direction (vanishing along $D$),
then one can make the $U(1)$ transformation $z\to e^{i\theta}z$.
This, roughly,
 is the symmetry whose possible anomaly we want to analyze.

Let us recall that the positive chirality spinor bundle $S_+$ of the
six-manifold $D$ has rank four.   The normal bundle
$\widehat N$ to $D$ in $\R^3\times X$ has  rank five, and
(since the spinor representation of $SO(5)$ has dimension four)
the spinor bundle $\tilde S$ constructed from $\widehat N$ has
rank four.  On the five-brane world-volume
$D$ propagate sixteen fermi fields $\psi$ transforming
as a section of $S_+\otimes \tilde S$.  
The world-volume action of the five-brane is not completely understood,
but the part quadratic in $\psi$ and not involving the two-form
is simply the Dirac action
for chiral spinors coupled to the metric of $D$ and to the $SO(5)$
structure group of  $\widehat N$:
\eqn\minac{L_f=\int_D d^6x\,\,\psi {\cal D}\psi.}
Here ${\cal D}$ is the Dirac operator coupled to the metric and $SO(5)$
gauge fields.  

For a generic six-dimensional submanifold
$D$ of an eleven-manifold, the Dirac action \minac\
has no global symmetries except mod two conservation of the number
of fermions.  We are here, however, in a special situation in which
 $\widehat N$ is simply  $N\oplus T\R^3$, where
$N$ is the normal bundle to $D$ in $X$ (now regarded as a rank
two real bundle), and $T\R^3$ is the (trivial) tangent bundle to $\R^3$.  
As a result, the $SO(5)$ structure group of the normal bundle
reduces to an $SO(2)$ (which acts only on $N$).  The subgroup
of $SO(5)$ that commutes with $SO(2)$ is $SO(2)\times SO(3)$ 
(the two factors correspond to rotations of $N$ and of $T\R^3$,
respectively), and $SO(2)\times SO(3)$ is therefore a symmetry
group of the classical action \minac.  We call the $SO(2)$ generator
$W$.  It is actually $W$ whose anomaly will control the generation
of a superpotential.

It is important to know whether $  W$ is an exact
symmetry of the five-brane action, or only a symmetry in the
approximation of \minac.  It is somewhat hard to answer this
question definitively because the five-brane action is not
fully known.  However, one may note that arbitrary couplings
of $\psi$ to fields defined on the five-brane will automatically
be $W$-invariant (with $W$ understood as acting trivially on
fields other than $\psi$).  To violate $ W$ would require
couplings of $\psi$ to the normal derivatives  (that is,
normal to $D$) of some
of the eleven-dimensional fields, for instance a coupling $\psi\psi R$
with $R$ the eleven-dimensional Riemann tensor.  It seems very likely
that such couplings are absent in the minimal supersymmetric
five-brane action, would vanish when $X$ is scaled up, and
would not contribute to the superpotential.  At any rate,
this assumption will be made in the present paper.

Because $\widehat N=N\oplus T\R^3$, the
spinor bundle 
$\tilde S$ of $\widehat N$ is  simply
the tensor product of a rank two spin bundle $S'$ derived
from $N$ with a constant rank two bundle $S''$ of spinors
of $T\R^3$.  The fermions on $D$ are thus simply two copies
of spinors with values in $S_+\otimes S'$, with the extra
two-valued index  (which comes from tensoring with
$S''$) transforming as spin one-half under rotations
of $\R^3$.  

Because of the relation of $N$ to the canonical bundle of $D$,
the spin bundle  $S'$ derived from $N$ is isomorphic
to $S'=K^{1/2}\oplus K^{-1/2}$, where $K^{1/2}$ is a square root
of $K$.  (It is not essential whether such a square root exists,
since the square roots will cancel out when we construct
$S_+\otimes S'$.) If we want
to keep track of the transformation law under $W$,
then we can write this as $S'=K^{1/2}_{1/2}\oplus K^{-1/2}_{-1/2}$,
where now the subscript  is the $W$ charge.

\def\O{{\cal O}}
\def\co{{\cal O}}
On the other hand, if $\Omega^{0,n}$ is the bundle of
complex-valued $(0,n)$-forms on $D$,   then (with an appropriate
matching of complex structure and orientation),
the positive and negative chirality spin bundles of $D$ are
\eqn\jurry{\eqalign{S_+&=K^{1/2}\oplus \left(
K^{1/2}\otimes \Omega^{0,2}\right)\cr
                    S_-&= \left(K^{1/2}\otimes \Omega^{0,1}\right)\oplus
                          \left(
K^{1/2}\otimes \Omega^{0,3}\right).\cr}}
(Rotations of the normal bundle act trivially on the tangent
bundle to $D$ and hence on $S_+$ and $S_-$.)
So the fermions on $D$ take values in 
\eqn\urry{S_+\otimes S'= \O_{-1/2} \oplus \Omega^{0,2}_{-1/2}
\oplus K_{1/2}\oplus \left(\Omega^{0,2}\otimes K_{1/2}\right).}
Here $\O$ (which is the same as $\Omega^{0,0}$) is a trivial
line bundle.

Now let $h_k$ be the dimension of the cohomology group 
$H^{0,k}(D)$, or equivalently the dimension of $H^{k,0}(D)$, the
space of holomorphic $k$-forms on $D$.  The number of fermion
zero modes with values in $\Omega^{0,n}$ is $h_n$, and
(by Serre duality) this also equals the number of zero modes
with values in $K\otimes \Omega^{0,3-n}$.  So, looking at \jurry,
we see that the number of $W=-1/2$ zero modes is $h_0+h_2$,
and the number of $W=1/2$ zero modes is $h_1+h_3$. Allowing
also for the doubling of the spectrum from tensoring with
the spinors of $T\R^3$, the total violation of $W$ because of the
fermion zero modes is
\def\co{{\cal O}}
\eqn\hurry{\Delta W = \chi(D,\co_D)=\sum_{n=0}^3 (-1)^nh_n,}
where $\chi(D,\co_D)$ is known as the arithmetic genus of $D$.
($\chi(D,\co_D)$ is sometimes abbreviated as $\chi(D)$, but this
notation can cause confusion with the topological Euler characteristic
of $D$.)  More generally, it   will be convenient, for any holomorphic
line bundle ${\cal L}$ on a complex manifold $ Y$, to define
\eqn\usefuldef{\chi(Y,{\cal L})=\sum_{i=0}^{{\rm dim}_\C\,Y}
(-1)^i{\rm dim}\,H^i(Y,{\cal L}).}
With $\co_D$ defined to be a trivial line bundle on $D$, the definition
\usefuldef\ for $Y=D$ and ${\cal L}=\co_D$ reduces to $\chi(D,\co_D)$
as defined in \hurry.

\nref\minasian{M. J. Duff, J. T. Liu, and R. Minasian,
``Eleven-Dimensional Origin Of String-String Duality: A One-Loop
Test,'' Nucl. Phys. {\bf B452} (1995) 261.}
\nref\five{E. Witten, ``Five-Branes And $M$-Theory On An Orbifold,''
hepth/9512219.}
Before proceeding, we might ask how to interpret this violation
of $W$.  After all, under a favorable condition
(if $X$ is the total space of a line bundle over a divisor $D$),
$W$ is simply the generator of a  diffeomorphism, and
the theory is supposed to be exactly invariant under diffeomorphisms!
The answer to this question \refs{\minasian,\five} is that, when
one interprets $W$ as generating a diffeomorphism,
 in addition to the violation of $W$ by
one-loop world-volume effects, there is also a ``classical'' violation
 that comes from a term $C\wedge I_8(R)$ in the low
energy expansion of $M$-theory; here $C$ is the massless three-form
and $I_8(R)$ is a homogeneous quartic polynomial in the Riemann tensor.
The $C\wedge I_8(R)$ term has a diffeomorphism anomaly that
cancels the one-loop anomaly of  the five-brane world-volume fields
for arbitrary diffeomorphisms and so in particular for $W$.  

Since the $C\wedge I_8(R)$ term is a ``classical'' effect, we should
expect to see it in the classical factor $e^{-(V_D+i\phi_D)}$ that was 
obtained above.  This factor should have an anomaly $-\chi(D,\co_D)$ 
under  
$W$, canceling the anomaly of the one-loop factor.  This must mean 
that under a rotation of the normal bundle, $\phi_D$ is shifted by a 
constant times $\chi(D,\co_D)$.  I will not derive this explicitly, but note 
that since the additive constant in $\phi_D$ is rather subtle to define,
there is room for such an effect.  

Now let us look at the fermion zero modes in this problem a little
more closely.
Two fermion zero modes are present universally.  These
are generated by the two supersymmetries that are unbroken in
compactification on $X$, but broken by wrapping a five-brane on $D$.
The two supersymmetries, being broken by the five-brane, generate
zero modes in the world-volume theory along $D$.  The wave function
of those zero modes is the constant
section 1 of $\O$. There are two of them once one tensors with the spinors
of $T\R^3$. 

Certain other fermi zero modes have a particularly simple interpretation.
A deformation of the complex divisor $D$ comes from a holomorphic
section of the normal bundle $N$, 
so the space of such deformations, in first order, is $H^0(D,N)$.
But the relation $N=K_D$ means that this space is just
$H^{3,0}(D)$.  So $h_3$ measures the number of possible first order
motions of the divisor $D$.

Now, here is a simple situation in which a superpotential {\it is}
generated: the case in which $h_1=h_2=h_3=0$.  
  The effect of the two fermion zero modes
that come from supersymmetry is simply that the classical
factor $e^{-(V_D+i\phi_D)}$ becomes a superpotential $\int d^2\theta
\,\,e^{-(V_D+i\phi_D)}$.  
Having $h_1=h_2=h_3=0$ means that there are no 
extra fermion zero modes that could cause the superpotential
to vanish. There are also, because $h_3=0$, no moduli in the position
of $D$; the presence of such moduli would be dangerous for
generating a superpotential, since the integration over the moduli
space might give a cancellation.\foot{When $h_3=0$, the only
bosonic moduli in the world-volume theory are the zero modes
of the self-dual two-form field, but these zero modes decouple
because of the two-form gauge invariance, so one gets no cancellation
from integration over the torus $H^2(X,\R)/H^2(X,\Z)$, to
which these modes are tangent.}
So when $h_1=h_2=h_3=0$, one gets a superpotential
interaction
\eqn\mikko{\int d^2\theta\,\, e^{-(V_D+i\phi_D)}T(m_\alpha),}
where $T(m_\alpha)$ is a holomorphic
function of other moduli $m_\alpha$ (such as the complex moduli
of $X$) which comes from the determinant for the {\it non-zero} modes
and so is  everywhere {\it non-zero}.

Now the factor $e^{-(V_D+i\phi_D)}$ carries charge
$ W=-\chi(D,\co_D)$, as we described earlier.  When $h_1=h_2=h_3=0$,
$\chi(D,\co_D)=1$, 
and in this case $e^{-(V_D+i\phi_D)}$ has $W=-1$.  It must
then be that the measure $d^2\theta$ has charge $W=1$, to make
it possible to generate the interaction \mikko\ under these
conditions.  In fact, the measure $d^2\theta$ {\it always}
carries $W=1$; this is clear from the fact that the supersymmetries
broken by the five-brane create the two ``universal'' fermion zero
modes that are sections of $\O_{-1/2}$.  While the measure
always has $W=1$, the factor $e^{-(V_D+i\phi_D)}$ has $W=-\chi(D,\co_D)$.
Therefore, {\it a superpotential can only be generated by 
wrapping a five-brane on $D$ if
$\chi(D,\co_D)=1$.}  We know a partial converse from the last paragraph:
 any divisor with $h_1=h_2=h_3=0$
{\it does} contribute to the superpotential. When
there are several such divisors, or when $h_3\not=0$ so that
there is a positive-dimensional moduli space of divisors to
integrate over, cancellations are conceivable.

By a standard argument in complex geometry, $\chi(D,\co_D)$ only
depends on the cohomology class of $D$.  Tautologically,
one defines a line bundle $\co_X(D)$ on $X$ that admits
 a holomorphic section
$s$ that vanishes precisely on $D$, and looks at the exact
sequence of sheaves
\eqn\nuggo{ 0 \to \O_X(-D) \to \co_X \to \co_D\to 0,}
where $\O_X$ and $\O_D$ are the trivial bundles (or ``structure
sheaves'') defined on $X$ and $D$, respectively, and $\O_X(-D)$ is the
inverse of $\O_X(D)$; the first
map in \nuggo\ is multiplication by $s$, and the second is restriction
to $D$.  The exact cohomology sequence derived from \nuggo\ then
implies that $\chi(D,\co_D)=-\chi(X,\O_X(-D)) +\chi(X,\O_X)$
(where $\chi(X,{\cal L})$ was defined in \usefuldef).
From the index theorem,
one deduces therefore that
\eqn\gluggo{\chi(D,\co_D) 
=\int_X\left(1-e^{-[D]} \right) \,\,{\rm Td}(X),}
where $ [D]=c_1(\O_X(D))$ is the cohomology class  dual to $D$,
and ${\rm Td}(X)$ is the Todd class.  This gives an explicit
formula for $\chi(D,\co_D)$ in terms of the cohomology class of
$D$, and severely limits the possible $ D$'s with $\chi(D,\co_D)=1$.

In the examples that follow, we either show that a superpotential
is not generated by instantons by
showing that any divisor $D$ on $X$ has
$\chi(D,\co_D)\not= 1$, or we show that a superpotential {\it is}
generated by showing that for some choice of the cohomology class there
is precisely one complex divisor $D$, which moreover has
$h_1=h_2=h_3=0$.  

In four-dimensional supersymmetric
field theory, it sometimes happens \ads\ that a superpotential
cannot be generated by instantons but is generated by non-perturbative
strong infrared dynamics.\foot{The author was reminded
of this by N. Seiberg.}  This apparently does not, however, happen in
$N=2$ theories in three dimensions, where (as the vector multiplet
contains a scalar that comes by dimensional reduction from a vector
in four dimensions) one can always go to a Coulomb branch with
the gauge group broken to an abelian subgroup, and turn off the
strong gauge dynamics.  So in such three-dimensional theories,
including the string theories that we have been studying, it
seems plausible that the instanton-induced superpotential is exact.

In section three, we  will consider partial decompactification
to four dimensions, but only in examples in which this occurs
without restoring a non-abelian gauge symmetry and producing
strong infrared dynamics.  Actually, in field theory, generation
of a superpotential by strong infrared dynamics (rather than
instantons) happens when symmetry violation
by instantons has the right sign but is too large to generate
a superpotential; then, sometimes, there is spontaneous symmetry
breaking that liberates effective ``fractional instantons'' with
the right quantum numbers.  In our problem, this might correspond 
to a situation in which there are divisors $D$ with $\chi(D,\co_D)$ 
positive,
but the smallest positive value is some $n>1$,
 so that to generate a superpotential one would formally need to wrap
$1/n$ five-branes over $D$.  We will not actually find such a situation
(possibly because the 
models we consider all have generically abelian
unbroken gauge groups even after decompactification to four
dimensions); in the models we consider, either a superpotential
is generated or $\chi(D,\co_D)$ is always negative.  In field theory,
when symmetry violation by instantons has the wrong sign, there
is no superpotential from instantons or otherwise.  I find it plausible
that that is also true in string theory. 

\subsec{Examples} 

For our first example, we consider a Calabi-Yau manifold
$X$ built as an intersection in $\P^{4+k}$
of $k$ hypersurfaces whose degree adds up to $5+k$.  Thus,
$X$ is defined by equations $g_1=\dots = g_k=0$, where the $g_j$
are homogeneous polynomials of degree $a_j$ in the homogeneous
coordinates of $\P^{4+k}$, and $\sum_{j=1}^k a_j=5+k$.  For instance,
we can take $k=1$ and consider a degree six equation in $\P^5$.

To find divisors in $X$, we first classify the possible line
bundles -- since every divisor $D$ is associated with a line
bundle $\O_X(D)$.  
Such line bundles are classified by their first Chern class in
$H^2(X,\Z)$.
To determine $H^2(X,\Z)$, one uses the Lefschetz theorem
(see for instance \nref\lefschetz{P. Griffiths and J. Harris,
{\it Principles Of Algebraic Geometry} (John Wiley and Sons, 1978),
p. 156.}
\nref\hubsch{T. Hubsch, {\it Calabi-Yau Manifolds: A Bestiary For
Physicists} (World-Scientific, 1992), p. 44.} \refs{\lefschetz,\hubsch}) 
which states that
if $Y$ is a complex manifold, and $s$ is a section of a positive
line bundle over $Y$, then the hypersurface $Z$ defined by $s=0$ has the
same (integral) cohomology as $Y$, up to the middle dimension.
(Thus, restriction from $Y$ to $Z$ gives an isomorphism from
$H^r(Y,\Z)$ to $H^r(Z,\Z)$ for $r<{\rm dim}_\C(Z)$; this isomorphism
is compatible with the Hodge decomposition.)  In our
problem, since the $g_j$ are all sections of positive line bundles,
repeated application of this theorem -- successively imposing
one after another of the equations $g_j=0$ --
implies that $H^2(X,\Z)$ coincides
(under restriction or pull-back) with $H^2(\P^{4+k},\Z)$, so that
the line bundles on $X$ are simply pull-backs of line bundles
on $\P^{4+k}$.

This means that the divisor $D$ is simply defined by an equation
$f=0$, with $f$ a homogeneous polynomial of some degree $n>0$ in
the homogeneous coordinates of $\P^{4+k}$.  $D$ is therefore defined
by the equations $f=g_1=\dots = g_k=0$ in $\P^{4+k}$.  Repeated
application of the Lefschetz theorem therefore implies that the
cohomology of $D$ coincides with that of $\P^{4+k}$ up to the middle
dimension, which is three, so that in particular
$h_j(D)=h_j(\P^{4+k})$ for 
$j=1,2$. Now $H^{n,m}(\P^r)=0$ except for $n=m$,
so that $h_n(\P^r)=0$ for $n>0$.  Hence $h_1(D)=h_2(D)=0$.
On the other hand, $h_3(D)$ is the number of complex deformations of
$D$ as a hypersurface in $X$, and this number, which is the number
of adjustable coefficients in the polynomial $f$, is strictly positive.
So $D$ has arithmetic genus less than one (and in fact negative).
Therefore, in $M$-theory compactification on $X$, no superpotential
is generated and supersymmetry is unbroken.

Now to give a simple example in which a superpotential {\it is} generated,
let $Y$ be any Calabi-Yau manifold with an isolated singularity
that looks locally like the quotient of $\C^4$ by the
$\Z_4$ group generated by $(x_1,x_2,x_3,x_4)\to (ix_1,ix_2,ix_3,ix_4)$.
Such a singularity can be resolved by blowing up the origin,
replacing it by a divisor $D$ that is a copy of $\P^3$.  Since
$h_n(\P^r)=0$ for $n>0$, such a divisor $D$ has $h_1=h_2=h_3=0$.
Moreover, any such divisor is the unique divisor in its cohomology
class, so cancellations are not possible.  
So on any smooth Calabi-Yau manifold $X$
obtained in this way, a superpotential is generated.  

For example, a singular Calabi-Yau manifold $Y$ with such orbifold
singularities can be constructed as a hypersurface of degree 12
in the weighted projective space $\P^5_{1,1,1,1,4,4}$, the subscripts
being the weights.  There are three $\Z_4$ orbifold singularities in $Y$,
at points at which the first four homogeneous coordinates vanish,
and their local structure is as described in the last paragraph.
So a superpotential is generated in compactification on the smooth
Calabi-Yau manifold $X$ obtained by blowing up these singularities.
($X$ can be constructed as a hypersurface in a $\P^2$ bundle
over $\P^3$, and so is related to other examples below.)

Now to 
consider a slightly more difficult example in which a superpotential
is {\it not} generated, consider the case of a hypersurface
$X$ of degree $(n+1,m+1)$ in $\P^n\times \P^m$, with $n+m=5$.  
Such a hypersurface is defined by an equation $g=0$, with $g$ being
a polynomial homogeneous of degree $n+1$ in the homogeneous coordinates
of $\P^n$ and of degree $m+1$ in the homogeneous coordinates of $\P^m$.
Since $g$ is a section of a positive line bundle over $\P^n\times\P^m$,
the Lefschetz theorem implies that any divisor $D$ on $X$ is given
by an equation $f=0$, where $f$ is a homogeneous polynomial of
degree $(a,b)$ in the homogeneous coordinates of $\P^n\times \P^m$,
for some integers $a,b$, which moreover must be non-negative (and not
both zero) for $D$ to exist.

If $a$ and $b$ are both positive, another application of the Lefschetz
theorem says that $h_1(D)=h_2(D)=0$.  Since $h_3(D)$ is positive
(equalling the number of variable parameters in $f$), such a divisor
does not contribute to the superpotential.  It remains to consider
the case that $a$ or $b$ is 0; there is no essential loss in generality
to suppose that $b=0$.  In this case, it is helpful to first look
at the hypersurface $Y$ defined by the equation $f=0$ in $\P^n$.
The Lefschetz theorem asserts that $Y$ has $h_1=h_2=0$ if $n\geq 4$,
and $h_1=0$ if $n=3$.  For $n=2$, $Y$ is a curve, which necessarily
has $h_2=0$, 
and for $n=1$, $Y$ is a finite set of points, with $h_1=h_2=0$.
Now, $Y\times \P^m$ has the same $h_j$ as $Y$, and the Lefschetz theorem
says that $D$ (which can be
defined by an equation $g=0$ in $Y\times \P^m$, where
$g$ is a section of a positive line bundle) has the same $h_1$ and $h_2$
as $Y\times \P^m$.  In particular, $h_2(D)=0$ unless $n=3$.  Since
also $h_3(D)>0$, if $h_2(D)=0$, $\chi(D,\co_D)<1$ 
and there is no contribution
to the superpotential.  

It remains then to look at the case $n=3$, 
for
which $h_1(D)=0$, but $h_2(D)$ and  $h_3(D)$ are both non-zero.
Explicitly, we are here dealing with 
a Calabi-Yau hypersurface $X$ in a product $\P^3\times \P^2$,
with homogeneous coordinates $(x_1,\dots ,x_4)$ and $(y_1,\dots, y_3)$,
respectively;
$X$ is defined by an equation $g=0$, with $g$ of degree $(4,3)$.
In the case not settled above, the divisor $D$ is given by
an equation $f(x_1,\dots,x_4)=0$, homogeneous of degree
$a>0$ in the $x_i$.  A holomorphic two-form on $D$ is of the
form
\eqn\giraffe{\omega_2= F(x_1,\dots,x_4){(x_1\,dx_2\wedge dx_3+
{\rm cyclic~permutations})\over \partial f/\partial x_4},}
with $F$ homogeneous of degree $n-4$.  Therefore, $h_2(D)$ is
the dimension of the space of polynomials that are homogeneous
of this degree.
On the other hand, a holomorphic three-form on $D$ is of
the form
\eqn\hiraffe{\omega_3=G(x_1,\dots , x_4){(x_1\,dx_2\wedge dx_3+
{\rm cyclic~permutations})\over \partial f/\partial x_4}{\left(
y_1\,dy_2-y_2dy_1\right) \over
{\partial g/\partial y_3}},}
with $G$ a polynomial homogeneous of degree $n$ (modulo
the relation $G\to G+\lambda g$, since $g =0 $ on $D$).  
Evidently, there are many more $G$'s than $F$'s,
so $0<h_2(D)<h_3(D)$, and in fact $\chi(D,\co_D)<0$.

Thus, there is no superpotential in compactification
on a Calabi-Yau hypersurface in $\P^n\times \P^m$.  Essentially
the 
same arguments can be used to show that there is also no superpotential
in compactification on a Calabi-Yau hypersurface in a product
$\P^{n_1}\times \dots \times \P^{n_k}$ of any number of projective
spaces.

There can, however, be a superpotential if one replaces
$\P^n\times\P^m$ by a $\P^m$ bundle over $\P^n$, still with $m+n=5$.
To describe such a bundle $Z$ for which there is a superpotential, 
introduce coordinates $x_1,\dots, x_{n+1}$
and $y_1,\dots, y_{m+1}$, and divide by $\C^*\times \C^*$ that acts
by
\eqn\pinkko{(x_1,\dots,x_{n+1},y_1,\dots,y_{m+1})
\to (\lambda x_1,\dots,\lambda x_{n+1},\mu y_1,\lambda\mu y_2,
\dots,\lambda\mu y_{m+1})}
with $\lambda,\mu\in \C^*$.
In other words, the $x_j$ transform with degree $(1,0)$ in $\lambda,\mu$,
while $y_1$ transforms with
 degree $(0,1)$, and the other $y_k$ with degree $(1,1)$.
Let $X$ be a Calabi-Yau manifold defined by an equation $g=0$ in $Z$,
 where $g$ is homogeneous of degree $(n+m+1,m+1)$.  I claim
that a superpotential is generated in compactification on $X$.

In fact, $y_1$ is the unique monomial of degree $(0,1)$, so the divisor
$D$ defined by $y_1=0$ is the unique divisor in its cohomology class.
Hence $h_3(D)=0$.  But since $D$ is defined by an equation $g=0$ in
$\P^n\times \P^{m-1}$, the Lefschetz theorem 
implies that $h_1(D)=h_2(D)=0$.
So a superpotential is generated.  An interesting feature of this
example is that the divisor $D$ on which the five-brane wraps to give
a superpotential is somewhat more general than in the previous examples.

One can similarly
work out other examples of hypersurfaces,
and intersections of hypersurfaces, in other toric varieties, for which
a superpotential is or is not generated.  It would be attractive to 
understand a systematic approach.   

\newsec{Application To $F$-Theory On A Calabi-Yau Four-Fold}

Now suppose that the Calabi-Yau four-fold $X$ can 
be elliptically fibered, that is that  there is a holomorphic
map $\pi:X\to B$ where
$B$ is a complex three-fold and the generic fibers are two-tori
$E$. Suppose moreover that $\pi$ has a holomorphic section.
\foot{One can actually proceed even if $\pi$ does not have such
a section; then $M$-theory on $X$ is equivalent to 
Type IIB on $\R^3\times \S^1\times B$ with non-zero
three-form field strengths on $\S^1\times B$, as explained below.
But I will here consider examples in which $\pi$ has a section
and the three-forms vanish in the Type IIB description.}
Then $M$-theory on $X$ is closely related \vafa\
to Type IIB superstring theory on $B$.  The relation is made
as follows.  Let $\epsilon$ be the area of $E$.  As $\epsilon\to 0$,
using fiber-wise the relation of $M$-theory on $\R^9\times \T^2$ with
Type IIB on $\R^9\times \S^1$, one replaces the two-torus
fibers $E$ with a fixed $\S^1$, also replacing $M$-theory with
Type IIB.  So $M$-theory on $\R^3\times X$
is Type IIB on $\R^3\times \S^1\times B$.  (This 
is  \vafa\ an unconventional
Type IIB perturbative vacuum; the Type IIB coupling varies with
the position on $B$.  Type IIB vacua of this kind are also called
$F$-theory vacua.) 
The radius of the $\S^1$ varies as an inverse power of $\epsilon$,
and so for $\epsilon\to 0$ one gets Type IIB on $\R^4\times B$.

We want to study superpotential generation by instantons in
Type IIB compactified on $B$.  This can be done simply by using
the conclusions of the last section, taking the limit as $\epsilon \to
0$.

For the present purposes, we should distinguish two kinds of
divisor $D$ on $X$, depending on whether the complex manifold
$\pi(D)$, which is a submanifold of $B$, is all of $B$ or a proper
submanifold.

$(a)$  In the first case, $D$ is either a ``section'' of $\pi$,
or a ``multisection'' (obtained by mapping holomorphically to $X$
a branched $m$-sheeted cover of $B$, for some $m>1$). 

$(b)$  Alternatively, one might take a divisor $C$ on $B$,
and set $D=\pi^{-1}(C)$ (or a component of $\pi^{-1}(C)$ in the
exceptional case in which $\pi^{-1}(C) $ has several components).

We will see that the divisors of type $(b)$  are the ones
that lead to instanton generation of a superpotential in
Type IIB compactification on $B$.  Let us start with an $M$-theory
five-brane wrapped on a divisor $D$ of type $(b)$ and see what it 
corresponds to in Type IIB theory on $B$.
Locally, when the fibers $E$ are small,  $\R^3\times X$
looks like $W\times \S^1\times \S^1$  where $E=\S^1\times\S^1$ and
$W$ is a nine-manifold; locally along $W$, 
the divisor $D$ is  $C\times \S^1\times \S^1$ where $C$ is a
four-cycle in $W$.
We first go to Type IIA by shrinking the second $\S^1$ in
$\S^1\times \S^1$.  We get
locally Type IIA on $W\times \S^1$, with the five-brane wrapped
on $D=C\times \S^1\times \S^1$ turning into a four-brane wrapped
on $C\times \S^1$.  Now we do $T$-duality on the remaining $\S^1$, going
over to Type IIB on $W\times \S^1$; this turns the four-brane
into a three-brane wrapped on $C$.  

Though the intermediate steps
here were local, the final answer holds globally:
 $M$-theory instantons of type $(b)$ correspond in $F$-theory
to three-branes whose four-dimensional world-volume
is wrapped on the divisor $C\subset B$.
The action for such an instanton is therefore of order $V_C$, the
volume of $C$.  In particular, this action is of order one if we take
$\epsilon\to 0$ while keeping the Type IIB or $F$-theory geometry fixed.

It is now evident that instantons of type $(a)$ do not survive
when we take $\epsilon\to 0$.
In fact, the volume of a divisor of
type $(a)$ is of order $1/\epsilon$ compared to that of a divisor
of type $(b)$ (if we take $\epsilon$ to 0 keeping fixed the geometry
of $B$), so divisors of type $(a)$ have action of order $1/\epsilon$
in $F$-theory units.

On the other hand, the divisors of type $(a)$ obviously have finite
action for non-zero $\epsilon$, and thus can contribute for
Type IIB on $\S^1\times B$ (which after all is the same as $M$-theory
on $X$).  

Now, as we will see presently,
there may or may not be a superpotential
in Type IIB compactification on $B$.  But there is always 
a superpotential (vanishing exponentially in the radius of the $\S^1$)
in Type IIB on $\S^1\times B$ (with vanishing three-forms).  
In fact, there is always a divisor
of type $(a)$, unique in its cohomology class, with $h_1=h_2=h_3=0$.
Such a divisor $D$ is the section of $\pi:X\to B$ which is part of the
defining data of $F$-theory.  Indeed, this $D$ is isomorphic to $B$,
but $B$ always has $h_1=h_2=h_3=0$.  The reason is that a holomorphic
$k$-form on $B$  would pull back under $\pi$  to a holomorphic
$k$-form on $X$; but the existence of such a holomorphic $k$-form
on $X$ for $k=1,2$, or 3 would contradict the Calabi-Yau property.
If the existence of a superpotential that vanishes when the
$\S^1$ becomes large is undesireable (as may be the case
\ref\ugh{E. Witten, ``Strong Coupling And The Cosmological
Constant,'' hepth/9506101.}), one can at least sometimes
eliminate it by turning on an $H$-field and replacing $X\to B$ with
a map that does not have a section, a situation discussed later.

In what follows, we consider only Type IIB compactification
to four dimensions on $B$, so we are interested only in divisors of type
$(b)$, that is, divisors $C$ in $B$.  To get a superpotential,
it is necessary for $C$ to  have the property that
$D=\pi^{-1}(C)$ has arithmetic genus 1, and sufficient to have
$h_j(D)=0$ for $j=1,2,3$.

\bigskip\noindent{\it Examples}

First, we consider some simple examples in which a superpotential
is not generated.  

We begin with $B=\P^3$.  A simple way to construct a Calabi-Yau
four-fold $X_0$ that is elliptically fibered over $\P^3$ is to take
 a hypersurface of degree $(4,3)$ in $\P^3\times \P^2$.
The map $\pi_0:X_0\to \P^3$ consists of forgetting the $\P^2$ factor;
the fibers are two-tori since a degree three equation in $\P^2$
defines a curve of genus one.
We considered this example in the last section 
in the context of $M$-theory and showed that no superpotential
is generated because every divisor $D$ has $\chi(D,\co_D)<1$.  
For $F$-theory (with $H=0$),
one cannot use $X_0$ because $\pi_0$ does not have a section.
Type IIB on $\P^3$ is constructed instead using a Calabi-Yau four-fold $X$
constructed as a hypersurface in a certain $\P^2$ bundle over $\P^3$
(and not simply in the product $\P^3\times \P^2$).  We call the total
space of this 
$\P^2$ bundle $W$ and let $\sigma:W\to \P^3$ be the projection.

We now must pick a divisor $C$ in $B=\P^3$, and compute the arithmetic
genus of $D=\sigma^{-1}(C)\cap X$.
To compute the invariants $h_j(D)$ requires a method
somewhat more powerful than used in the last section.
 One can conveniently use
the spectral sequence for the projection $\sigma: \sigma^{-1}(C) 
\to C$ 
as in Proposition 2.2 in \ref\grassi{A. Grassi, ``On Minimal Models Of
Elliptic Threefolds,'' Math. Ann. {\bf 290} (1991) 287.}.\foot{This
reference was pointed out by M. Gross, who also showed  that the
condition on normal crossings of the discriminant can be replaced by
the fact that $D$ lies in a $\P^2$ bundle over $C$.}  
The conclusion
is   similar to what we found in the last section for a divisor
of degree $(a,0)$ in $X_0$:
$h_1(D)=0$, $0<h_2(D)<h_3(D)$, so $\chi(D,\co_D)<1$.  In fact, in using
Proposition 2.2 of \grassi, it does not matter whether the $\P^2$ bundle
over $\P^3$ is trivial or not; in either case, one gets an
explicit description
of holomorphic forms on $D$ along the lines of \giraffe\ and \hiraffe,
with an obvious counting showing that $h_2(D)<h_3(D)$.
So there is no instanton-generated
superpotential in $F$-theory on $\P^3$.  (The fact that in the
computation it does
not matter whether the map $X\to \P^3$ has a section has
a physical explanation given below.) 

The same conclusion can be reached in the same way if $\P^3$ is
replaced by $\P^2\times\P^1$ or $(\P^1)^3$, or, roughly,
any example in which the normal bundle to a divisor always has enough
positivity.   These latter examples
have some interest because (by forgetting one of the $\P^1$ factors)
they are fibered over $\P^2$ or $\P^1\times \P^1$, with fiber $\P^1$;
they thus have interpretations in terms of the heterotic string,
as we discuss in the next section.

For a rather different example, suppose that $B$ is obtained from
another surface, such as $\P^3$, by blowing up a point $x$, an
operation that replaces $x$ by a divisor $C$ isomorphic to $\P^2$,
with normal bundle ${\cal O}(-1)$.  Then given an elliptic fibration
$\pi:X\to B$, let $D=\pi^{-1}(C)$.  One can compute, for example
by again using Proposition 2.2 of \grassi, that $h_j(D)=0$ for
$j=1,2,3$, so that the wrapping of a Type IIB three-brane over
such an ``exceptional divisor'' $D$ does generate a superpotential.
\foot{There is actually a puzzle here, because one would expect
as in \fm\ and the second paper in \morrisonvafa\ to see a  phase
transition from Type IIB on $\P^3$ with a point blown up to Type IIB
on $\P^3$; this seems to be a transition from a phase with
broken supersymmetry to a phase with unbroken supersymmetry, something
that one would not usually expect.} 

This last example has an interpretation in terms of the heterotic string,
since $\P^3$ with a point blown up can be fibered over $\P^2$ with
fibers $\P^1$.  To see this, consider $\C^5$ with coordinates
$(x_1,x_2,x_3,u,v)$.   Define a three-fold $B$ to be the
quotient of $\C^5$ (with the points with $u=v=0$ or $x_1=x_2=x_3=u=0$
 deleted) by
a  $\C^*\times \C^*$ action
$(x_1,x_2,x_3,u,v)\to(\lambda x_1,\lambda x_2,\lambda x_3, \mu u,
\lambda\mu v)$.    
Then $B$ is fibered over $\P^2$ by forgetting $u,v$; the
fibers are $\P^1$'s, obtained by projectivizing $u,v$.  The divisor
$C$ with $u=0$ is a section of the fibration $B\to \P^2$.  
$C$ is isomorphic
to $\P^2$; once $u$ is set to zero, the scaling by $\mu$ can be used
to eliminate $v$, and then $x_1,x_2,x_3$ are interpreted as homogeneous
coordinates for $C\cong \P^2$.  The normal bundle to $C$ in $B$ is
${\cal O}(-1)$ (in scaling by $\lambda$, 
to preserve a ``gauge condition'' $v=1$ by which
$\mu $ and $v$ were eliminated, one sets $\mu=\lambda^{-1}$, so that
the normal coordinate $u$ to $C$ scales as $\lambda^{-1}$).
One maps $B$ to $\P^3$ by $(x_1,x_2,x_3,u,v)\to (ux_1,ux_2,ux_3,v)$.
This is an isomorphism away from $u=0$, and 
 ``blows down'' the divisor $C$ to the point $(0,0,0,1)$.

Clearly, the three-fold $B$ just considered is rather similar to the
Hirzebruch surface ${\bf F}_1$, used in Type IIB compactification
in \morrisonvafa.  This surface is isomorphic to $\P^2$ with
a point blown up, or to a $\P^1$ bundle over $\P^1$, with a section,
$E$, of self-intersection $-1$.
For our final example, take  $B=G\times {\bf F}_1$, with $G$ another
copy of $\P^1$.  Note that $B$ can be given two different structures
of $\P^1$ fibration: one has $\tau:B\to {\bf F}_1$ by forgetting $G$,
or $\tau':B\to G\times \P^1$ by taking the product of the identity
map on $G$ with the projection ${\bf F}_1\to \P^1$.
This will lead to two different identifications with the heterotic
string rather as the existence of two different K3 fibrations
has been exploited \nref\aspinwallgross{P. Aspinwall and M. Gross,
``Heterotic-Heterotic String Duality And
Multiple K3 Fibrations,'' hepth/9602118.}
\refs{\aspinwallgross,\morrisonvafa}.

Now, let $C$ be the divisor $G\times E$ in $G\times {\bf F}_1$.  Let
$X$ be a Calabi-Yau four-fold with an elliptic fibration $\sigma:X\to B$.
Using proposition 2.2 of \grassi, the divisor $D=\sigma^{-1}(C)$ in $X$
can be shown to have 
$h_j=0$, $j=1,2,3$, so therefore a superpotential is generated
in wrapping a Type IIB three-brane on $C$.  Note that $C$ is a section of
$\tau'$, but not of $\tau$; rather,
 $C=\tau^{-1}(C')$ where $C'=E$ is a curve in $\tau(B)={\bf F}_1$.
The consequences of these statements for the heterotic string
will be clear in the next section.

\bigskip\noindent
{\it Elliptic Fibrations Without A Section}

Finally, let us briefly consider $M$-theory compactification
on an elliptically fibered four-fold $X$ that does {\it not}
have a section.

As the fibers shrink, $\R^3\times X$ looks locally
like $W\times \S^1\times \S^1$ with a nine-manifold $ W$.  Absence
of a section means that, calling the last two coordinates
$x^{10}$ and $x^{11}$, there are terms in the metric $g_{i\,10}$ and
$g_{i\,11}$ (with $i=1,\dots,9$) 
that cannot be eliminated by shifting $x^{10}$ and $x^{11}$
by functions of the first nine coordinates.  After shrinking
the second circle, one locally along $W$ gets Type IIA on $W\times
\S^1$ with a non-trivial $g_{i\,10}$; the $T$-duality that
maps this to Type IIB on $W\times \S^1$ (which is the description
that makes sense globally) turns this into a non-zero $B_{i\,10}$,
with $B$ the massless Neveu-Schwarz
two-form of the theory; absence of a section
in the original description means that $H=dB$ is non-zero.  The
Ramond two-form $\tilde B$ of the Type IIB theory is also non-zero,
since it arises from $g_{i\,11}$ in the same chain of dualities.   
Of course, they are related to each other by $SL(2,\Z)$, so in
$F$-theory one could not have one without the other.

The $H$-fields obtained this way have topologically normalized
periods, and therefore
vanish as forms in the limit as the radius of the $\S^1$
is scaled up and one goes to four dimensions.  This gives a physical
explanation for the fact (which was exhibited above as a consequence
of Proposition 2.2 of \grassi) that to analyze the part of the
superpotential that     survives when one gets to four dimensions,
it does not matter whether the morphism $X\to B$ has a section.

\newsec{Application To The Heterotic String On A Calabi-Yau Three-Fold}

According to \vafa, Type IIB on $\P^1$ is the same as the heterotic
string on $\T^2$.  Therefore, if the complex three-fold $B$ is
fibered over a two-fold $B'$ with fibers $\P^1$, by a holomorphic
map $\tau:B\to B'$ with $\P^1$ fibers, then Type IIB on $B$ is
equivalent to the heterotic string on a Calabi-Yau three-fold
$Z$ that is fibered over $B'$ with the $\P^1$ fibers replaced
by $\T^2$'s.  For the  analogous case with 
$B$ and $B'$  of complex dimension
two and one, this construction has been used \morrisonvafa\ 
to study the heterotic string on K3.

The equivalence of certain heterotic string models to Type IIB 
compactifications makes it possible to control the superpotential
by studying divisors, as in the last section.
Some of the examples given in the last section admit such $\P^1$
fibrations $B\to B'$, and we need not repeat the examples here.
I will not try in this paper to be explicit about the precise heterotic
string models that these Type IIB compactifications correspond to.
(Many relevant facts are in the second paper in \morrisonvafa.)
But I will compare the qualitative results to what is expected 
of heterotic string physics.

We classify divisors $C\subset B$ according to whether $\tau(C)$ is
all of $B'$ or a submanifold $C'\subset B'$:

$(a')$ In the first case, $C$ is a section or multi-section of
$\tau:B\to C$.

$(b')$ In the second case, we start with a Riemann surface $C'\subset B'$,
and $C=\tau^{-1}(C')$ (or possibly a component thereof).

Unlike the corresponding situation in $F$-theory, divisors of either
type may contribute to the superpotential, since we are not interested
in taking any particular limit on the area of the $\P^1$.
In the heterotic string, one expects 
at least two weak coupling mechanisms for generating
a superpotential, namely:

$(a'')$ Space-time instantons.

$(b'')$ World-sheet instantons.

I claim that contributions from
divisors of type $(a')$ correspond to space-time instanton effects,
and contributions from divisors of type $(b')$ 
correspond to world-sheet instanton effects.  

A preliminary check is as follows.
Note that space-time instantons on $\R^4\times Z$ are localized in
$\R^4$ but spread out over $Z$.  
By contrast, world-sheet instantons are localized on a Riemann surface
$F\subset Z$.  In the Type IIB description
on $\R^4\times B$, we cannot conveniently see 
$Z$, but we can conveniently see the four-dimensional space $B'$ that
$Z$ maps to.  It is clear that divisors of type $(a')$, which map to all
of $B'$, cannot be localized on a submanifold of $Z$ of real dimension less
than four, and so could not correspond to world-sheet instantons.
However, divisors of type $(b')$ are localized in two dimensions on $B'$,
and so might possibly be localized on a two-dimensional submanifold
of $Z$ and correspond to world-sheet instantons.

For more precise information, begin with Type IIB theory compactified
to eight dimensions on a $\P^1$ of volume $V$.  The action for
the massless graviton  and gauge fields on $\R^8$  is qualitatively
\eqn\jurgle{L=\int_{\R^8} d^8x\sqrt g \left(V R+ \tr F^2\right),}
with $g$ the metric in the Type IIB description, 
$R$ the Ricci scalar, and $F$ the Yang-Mills
field strength.  
(There is no dilaton in the formula since the coupling varies on $\P^1$
in a way uniquely determined by the vector moduli, which have been
 suppressed.)
The point here is that the gravitational action has a factor of $V$ 
from integration over $\P^1$, but the gauge fields are supported 
at special points on $\P^1$ (related to singularities of the $F$-theory
fibration) and have no such factor of $V$.  To go to a heterotic
string description, one introduces the heterotic string metric $g_h$ by
$g=V^{-1}g_h$, whereupon one gets
\eqn\urgle{L=\int_{\R^8}d^8x\sqrt{g_h}V^{-2}\left(R_h+\tr F^2\right).}
($R_h$ is the Ricci scalar constructed from $g_h$.)  From \urgle\ we
see that the heterotic string coupling is $\lambda_h=V$ \vafa.

Note that in the heterotic string description, the $\P^1$ is replaced
by a $\T^2$ whose volume, at a generic point in
Narain moduli space, is of order
\eqn\ollo{V_h=1.}

Now we want to consider what happens when one compactifies to four
dimensions on a manifold $B$ with a $\P^1$ fibration.  For simplicity,
we consider a product $B=\P^1\times B'$, rather than a fibration.
Let $V'$ be the volume of the four-manifold $B'$, in Type IIB units.
Because of the relation $g=V^{-1}g_h$, the volume of $B'$ in heterotic
string units is then $V'_h=V^2V'$.  
Because of \ollo, the volume of $V_Z$ of $Z$ in heterotic string units
is also of order $V'_h$.  The action $I_{ST}$  
of a space-time instanton of the
heterotic string is of order $V_Z/\lambda_h^2$; combining the above
formulas we get 
\eqn\umbo{I_{ST}= V'.}
But $V'$ is the action, in the Type IIB description, of an instanton
of type $(a')$, which corresponds simply to a three-brane that wraps
over $B'$, whose volume is $V'$. 

Now consider a divisor of type $(b')$, coming from a Riemann surface
$C'\subset B'$ whose area in Type IIB units is $A$.  The volume of the divisor
$B'=\tau^{-1}(C')$ is then  
\eqn\jumbo{{\rm Vol}(B') = VA,}
and this is the action in the Type IIB description of a three-brane
wrapping over this divisor.  In the heterotic string description,
if an instanton of type $(b')$ is going to correspond to a world-sheet
instanton, then this instanton will have to correspond to a Riemann
surface $C''\subset Z$ which is mapped to $C'$ by
the projection $Z\to B'$.  The action $I_{WS}$ of
such an instanton is of order the area $A_h$ of $C'$ in the 
heterotic string
description; because of the Weyl transformation between the two metrics,
$A_h=VA$.  So comparing to \jumbo, we get the desired relation
\eqn\bumbo{I_{WS}={\rm Vol }(B').} 

From the examples of section (3), one sees instantons of either kind
contributing to heterotic string superpotentials.  Curiously, as was
explained in the introduction, there is some novelty in this even
for the world-sheet instantons.
\listrefs
\end